\documentstyle[epsfig]{cjp3}
\begin{document}
\title{Topics in  non-perturbative QCD}
\authori{Adriano Di Giacomo}
\addressi{Dipartimento di Fisica E. Fermi, Universit\`a di Pisa, and
I.N.F.N.
    Via Buonarroti 2, 56100 Pisa}
\authorii{}       
\addressii{}
\authoriii{}      
\addressiii{}     



\headtitle{Topics in  non-perturbative QCD}
\headauthor{Adriano Di Giacomo}



\evidence{A}

\daterec{XXX}    

\cislo{0}  \year{2000}

\setcounter{page}{1}



\maketitle

\section{QCD. Perturbative and non perturbative.}
\subsection{Basic field theory.}
A quantum system is defined by the canonical variables $q(t)$, $p(t)$, and
by the Hamiltonian $H(p,q)$.

$q(t)$, $p(t)$ is a short notation for $q_i(t)$, $p_i(t)$, the index $i$
running over the degrees of freedom. In a field theory
\begin{equation}
q_i(t) = \varphi_a(\vec x,t)\label{eq1.1}
\end{equation}
or $i\equiv (a,\vec x)$. There are infinitely many degrees of freedom.

Solving the system means to construct a Hilbert space on which $q$, $p$
act as operators obeying canonical commutation relations at equal time
and the equations of motion. A ground state must exist.

The usual approach in quantum field theory is to split the Hamiltonian
$H$ in a free Hamiltonian $H_0$ and an interaction $H_I$
\begin{equation}
H = H_0 + H_I\label{eq1.2}
\end{equation}
and to assume as Hilbert space the Hilbert space of the free system
$H_0$, and as a ground state the corresponding Fock vacuum. Scattering
amplitudes between fundamental excitations are then computed by
perturbation theory. This approach is known as perturbative quantization,
and is what is presented in textbooks.

It is imediately apparent that $QCD$ will present problems in this
procedure: free quarks and gluons are not observed in nature. Perturbing
in their Fock space does not seem a sensible procedure.

A more fundamental approach is needed.
\subsection{Feynman-path formulation.\cite{r1}}
The basic quantity in Feynman's formulation is the action
\begin{equation}
S[q] = \int dt {\cal L}[q,\dot q]\label{eq1.3}
\end{equation}
where ${\cal L} = p\dot q - H$ is the Lagrangian.

$S[q]$ is a functional of $q(t)$.

In terms of $S$ the equations of motion are
\begin{equation}
\frac{\displaystyle\delta S}{\displaystyle\delta q_i(t)} = 0\label{eq1.4}
\end{equation}
The symbol in eq.(\ref{eq1.4}) means functional derivative.

The solution of the system is equivalent to compute the partition function
$Z$
\begin{equation}
Z = \int\left[\prod_{i,t} dq_i(t)\right]{\rm exp}\{i S[q]\}
\label{eq1.5}
\end{equation}
or better the generating functional
\begin{equation}
Z[J] = \int\left[\prod_{i,t} dq_i(t)\right]{\rm exp}\{i S[q]+
i\int dt J_i(t) q_i(t)\}
\label{eq1.6}
\end{equation}
with $J_i(t)$ arbitrary functions of time.

The integrals in (\ref{eq1.5}), (\ref{eq1.6}) are on a continuous
infinity of variables (functional integrals).

$Z[J]$, if it exists, provides the solution of the system, in the sense
that all correlation functions of the fields are known in terms of $Z[J]$
\begin{equation}
\frac{1}{Z}
(-i)^n\left. \frac{\displaystyle \delta^n Z[J]}
{\displaystyle\delta J_{i_1}(t_1) \ldots
\delta J_{i_n}(t_n)}\right|_{J=0} =
\langle 0| T\left(
q_{i_1}(t_1) \ldots q_{i_n}(t_n)\right)|0\rangle
\label{eq1.7}
\end{equation}
The knowledge of the correlation functions is indeed equivalent to the
knowledge of the Hilbert space, field operators and ground state
(reconstruction theorem).

The proof of (\ref{eq1.7}) goes as follows.

The correlations functions obey the following set of coupled differential
equations, with boundary conditions fixed by the $T$ prescription
\begin{eqnarray}
&&{\cal D}_t
\langle 0| T\left(q(t)
q_{i_1}(t_1) \ldots q_{i_n}(t_n)\right)|0\rangle\label{eq1.8}\\
&=&\sum_{k=1}^n
\delta(t - t_{k})\delta_{i i_k}
\langle 0| T\left(q(t)
q_{i_1}(t_1) \ldots
q_{i_{k-1}}(t_{k-1})q_{i_{k+1}}(t_{k+1})
q_{i_n}(t_n)\right)|0\rangle
\nonumber
\end{eqnarray}
Here
\begin{equation}
{\cal D}_t q(t) = 0 \label{eq1.9}
\end{equation}
indicate the equations of motion, and putting ${\cal D}_t$ out  of the
$T$ product means that time derivatives contained in ${\cal D}_t$
must act on the $T$ product. By use of eq.(\ref{eq1.9}) the only terms
which survive are those obtained by differentiating the $\theta$
functions in the $T$ product, giving the $\delta$ functions in the right
hand side, and canonical equal time commutators.

The left hand side of eq.(\ref{eq1.7}) has the $T$ prescription built in,
and obeys the same differential equations, which proves the equality.

Indeed
\begin{eqnarray}
&&\frac{1}{Z}
\int[\prod d q] {\cal D}_t q_i(t)
q_{i_1}(t_1) \ldots q_{i_n}(t_n) {\rm exp}[-i S(q)]=\nonumber\\
&=&
\frac{1}{Z}
\int[\prod d q] \frac{\delta S}{\displaystyle \delta q_i(t)}
q_{i_1}(t_1) \ldots q_{i_n}(t_n) {\rm exp}[-i S(q)]=\nonumber\\
&=&
\frac{i}{Z}
\int[\prod d q]
\left\{
\frac{\delta }{\displaystyle \delta q_i(t)} {\rm exp}[-i S(q)]\right\}
q_{i_1}(t_1) \ldots q_{i_n}(t_n) = \label{eq1.10}\\
&=&
\frac{i}{Z} \int[\prod d q]
\sum_{k=1}^n
\delta(t - t_{k})\delta_{i i_k}
{\rm exp}[-i S(q)]
q_{i_1}(t_1) \ldots
q_{i_{k-1}}(t_{k-1})q_{i_{k+1}}(t_{k+1})
q_{i_n}(t_n)\nonumber
\end{eqnarray}
The first equality follows from eq.(\ref{eq1.4}), the third by partial
integration. The finite term drops if ${\rm exp}[-i S(q)]$ vanishes at
large $q$'s. In the euclidean this amounts to have $S(q)\to\+\infty$ as
$q\to \pm\infty$ (stability). The argument is not true for a $\varphi^3$
theory or if in a $\varphi^4$ theory the coefficient of $\varphi^4$ has
the wrong sign.

A more direct proof is the original construction of Feynman\cite{r1}. $q$
is  a complete set of operators. The amplitude
\begin{equation}
{}_{t'}\langle q' | q\rangle_t =
\langle q' |{\rm exp}[-i H(t'-t)] q\rangle\label{eq1.11}
\end{equation}
contains all the physics of the system.

We now split the time $t'-t$ into $n+1$ intervals of length $\delta$,
with the idea of sending $n\to\infty$
\[\delta = \frac{t'-t}{n+1}\, \mathop{\to}_{n\to\infty}\, 0\]
Then, using completeness
\begin{equation}
{}_{t'}\langle q' | q\rangle_t =
\int dq_1\ldots dq_n
\langle q' | e^{-i H \delta} | q_n\rangle
\langle q_n|e^{-i H \delta} | q_{n-1}\rangle\ldots
\langle q_1|e^{-i H \delta} | q\rangle
\label{eq1.12}
\end{equation}
Consider, for the sake of simplicity, $p$ independent forces
\[ H = \frac{p^2}{2} + V(q)\]
Then
\begin{eqnarray}
&&\langle q_{k+1}|e^{-i H \delta} | q_k\rangle
\mathop{\simeq}_{{\cal O}(\delta^2)}
e^{-i V(q) \delta}
\langle q_{k+1}|
e^{-i \frac{1}{2} p^2 \delta} | q_k\rangle
=\nonumber\\
&&{\rm exp}\left\{i\delta\left[\frac{\displaystyle(q_{k+1}-q_k)^2}
{\displaystyle 2\delta^2} - V(q)\right]\right\}
={\rm exp}\left\{i\delta {\cal L}[q_k]\right\}
\label{eq1.13}
\end{eqnarray}
and
\begin{equation}
{}_{t'}\langle q' | q\rangle_t =\lim_{n\to\infty}
\int\prod_1^n d q_k e^{i S_n(q)}\label{eq1.14}
\end{equation}
Eq.(\ref{eq1.14}) is the definition of a functional integral as a limit
of ordinary integrals.

The above construction also provides a continuation to Euclidean time.

Consider the amplitude
${}_{iT}\langle q' | q\rangle_{-i T}$ as $T\to\infty$, in the presence
of an external source $J(\tau)$ which is non zero in the interval $-T/2$,
$T/2$. Let $| E_n\rangle$ be a complete set of states of given energies
\begin{eqnarray}
\lim_{T\to\infty}{}_{iT}\langle q' | q\rangle_{-i T}
&=&\hskip-10pt
\sum_{E_n, E_{n'}}
\langle q'| e^{-HT/2}| E_{n'}\rangle
\langle E_{n'}| \int\left[\prod dq\right]\;
e^{-S[q] - \int J(t) q(t) dt} | E_n\rangle\langle E_n
|  e^{-HT/2}| q\rangle\nonumber\\
&\simeq&
\langle q'|0\rangle\langle 0|q\rangle
\langle0| \int\left[\prod dq\right]\;
e^{-S[q] - \int J(t) q(t) dt} | 0\rangle\label{eq1.15}
\end{eqnarray}
Only the ground state survives the limit.
\subsection{Feynman integral and perturbation theory.}
To describe perturbation theory in the language of Feynman integral, the
Lagrangian is split into a term bilinear in the fields ${\cal L}_0$ plus
the rest
\[{\cal L} = {\cal L}_0 + {\cal L}_I\]
or
\begin{equation}
S[q] = S_0 + S_I\label{eq1.16}
\end{equation}
where, in generic notation
\begin{equation}
S_0 = \frac{1}{2} q_\alpha M^{-1}_{\alpha\beta} q_\beta\label{eq1.17}
\end{equation}
and $S_I$ is of higher degree in the fields. The matrix in
eq.(\ref{eq1.17})is denoted $M^{-1}$ for convenience.

Then
\begin{equation}
{\rm exp}[-S] = {\rm exp}(-S_0)
\left\{\sum_n \frac{(-1)^n}{n!} S_I^n\right\}\label{eq1.18}
\end{equation}
A power exansion is performed in $S_I$. The partition function becomes
\begin{equation}
Z = \int\left[ \prod dq\right]
{\rm exp}(-S_0)
\left\{\sum_n \frac{(-1)^n}{n!} S_I^n\right\}\label{eq1.19}
\end{equation}
and
\begin{equation}
\langle 0| T\left(
q_{i_1}(t_1) \ldots q_{i_n}(t_n)\right)|0\rangle
=\frac{1}{Z}
\int\left[ \prod dq\right]
e^{-S_0} q_{i_1}(t_1) \ldots q_{i_n}(t_n)
\left[\sum_n \frac{(-1)^n}{n!} S_I^n\right]
\label{eq1.20}
\end{equation}
If $S_I$ is a polynomial in the fields the generic integral in
(\ref{eq1.20}) is gaussian, i.e. has the form
\begin{equation}
\int\prod dq_i\,
e^{-\frac{1}{2}q_\alpha M^{-1}_{\alpha\beta} q_\beta}
q_{\alpha_1}\ldots q_{\alpha_n} =
\sum\prod_{\left\{\alpha_i,\alpha_j\right\}} M_{\alpha_i\alpha_j}
\label{eq1.21}
\end{equation}
The sum is extended to all possible set of pairs of the indices, $M$ is
the propagator, and (\ref{eq1.21}) is nothing but Wick's theorem.

A technical point arises in gauge theories where the matrix $M^{-1}$ is
not invertible, and the naive propagator does not exist due to gauge
invariance: the way around is Faddeev-Popov procedure\cite{r2}, which
brings back to
eq.(\ref{eq1.21}), with a well defined $M$, and the addition of ghost
fields.

The procedure explained above, to expand the weight function in $Z$ in
powers of $S_I$ is far from harmless. This is physically expected,
because perturbing around the Fock vacuum of quarks and gluons and
computing scattering amplitudes thereof, is not a good approximation in a
theory which confines them.

This desease manifests itself in the fact that renormalized expansion of a
generic observable ${\cal O}$ in powers of the coupling constant
$\alpha_s = g^2/4\pi$
\begin{equation}
{\cal O} \sum \alpha_s^r {\cal O}_r\label{eq1.22}
\end{equation}
is not convergent, not even as an asymptotic series\cite{r3}.
\subsection{Feynman path and lattice QCD.}
The lattice formulation\cite{r4} is a discrete approximant of QCD Feynman
integral, in the sense of eq.'s (\ref{eq1.12}), (\ref{eq1.14}). The
approximant is expected to be good if the lattice is large compared to the
correlation length $\lambda$ and $\lambda \gg a$, the lattice spacing.
The building block of the theory is the parallel transport between two
neighbouring sites of the lattice
\begin{equation}
U_\mu(\vec n) = {\rm exp}\left[i a g A_\mu(\vec n)\right]\label{eq1.23}
\end{equation}
where $\vec n$ is the site, $\mu = 1\ldots 4$ the direction to the next
site, and
\[ A_\mu = \sum^a T^a A_\mu^a\]
with $T^a$ the generators of the group in the fundamental representation.
The action is related to the parallel transport around the elementary
square surface $\Pi_{\mu\nu}$ (plaquette)
\begin{equation}
S = \beta\,\sum_{n,\mu\neq\nu}{\rm Tr}\left\{1 - \Pi_{\mu\nu}\right\}
\qquad \beta = \frac{\displaystyle2 N_c}{g^2}
\label{eq1.24}
\end{equation}
In the limit $a\to 0$
\begin{equation}
S \mathop{\simeq}_{a\to 0}
-\frac{a^4}{4}\sum_{n,\mu,\nu} G^a_{\mu\nu}G^a_{\mu\nu} + {\cal O}(a^6)
\label{eq1.25}
\end{equation}
i.e. $S$ differs from the continuum action by ``irrelevant'' terms.

It can be shown that the lattice action defines a theory with an
hermitian Hamiltonian, or that, in the language of statistical mechanics,
a transfer matrix exists.

The theory has a mass gap: the string tension $\sigma$ is observed in
numerical simulations.

The theory has a fixed point as $\beta\to\infty$, where the correlation
length diverges.

In formulae the lattice spacing is related to the physical scale
$\Lambda$ as
\begin{equation}
a = \frac{1}{\Lambda} {\rm exp}(- b_0\beta)\label{eq1.26}
\end{equation}
where $-b_0 < 0$ is the first coefficient of the perturbative $\beta$
function.

Eq.(\ref{eq1.26}) means that as the critical point $\beta=\infty$ is
approached the lattice spacing becomes exponentially small in physical
units, and the coarse structure of the lattice disappears (scaling
region).

In summary if $\beta$ is sufficiently large, so that the physical
correlation length $\lambda$ is such that $\lambda \gg a$ and at the same
time $\lambda \ll L$, the size of the lattice,the lattice should provide
a good approximant to continuum Feynman integral.

The partition function
\begin{equation}
Z = \int\left[\prod d U_\mu(n)\right] {\rm exp}[- S]
\label{eq1.27}
\end{equation}
is finite, since the group is compact.

The formulation is gauge invariant and needs no Faddeev Popov.

Of course it would be perfect if anybody could compute $Z$ analitically,
and then perform the continuum limit. The problem is open for competition.

A practical approach\cite{r5} is to compute $Z$ numerically, by producing
via Montecarlo techniques a ``significant sample'' of configurations on
which expectation values can be computed. As we shall see below if
properly asked lattice can give important answers.
\subsection{Non convergence of perturbative expansion. Non perturbative
formulation of QCD.}
As first noticed by Dyson\cite{r6} QED cannot be analytic in $\alpha$ at
$\alpha=0$. A gas of $N$ electrons in a given volume has energy
\[ E \propto c N + d \alpha N^2\]
with $c,d$ positive coefficients. The second term stems from Coulomb
repulsion. If $\alpha\to-\alpha$, the system becomes unstable, and hence
no neighbour of the point $\alpha=0$ exists in which physics is analytic
in $\alpha$.

Still QED works as an effective theory.

As we will see below a similar phenomenon occurs in QCD. Perturbation
theory is ill-defined. Still it works at short distances, for reasons
which are not really understood. In addition QCD, as argued in sect.4, is
not an effective theory, but exists in the sense of constructive field
theory.

Lattice is the most rigorous approach beyond perturbation theory. There
exist other attempts in the same direction, which we will briefly comment
upon
\begin{enumerate}
\item A phenomenological approach is the SVZ sum rules\cite{r7}: non
perturbative effects are parametrized by condensates, which are defined
via Wilson Operator Product Expansion (see sect.~2.1 below).
\item An interesting approach is known as $N_c\to\infty$\cite{r8}. The
idea is that the structure of QCD is almost independent on the number of
colours
$N_c$: the theory at $N_c=\infty$ differs from that at finite $N_c$ by
small corrections ${\cal O}(1/N_c)$ which do not alter the main physical
features. More precisely the limit is defined as $N_c\to\infty$ at
$g^2N_c=\lambda$ fixed. Only planar graphs survive at $N_c=\infty$.
Quarks loops are non leading in this expansion. Indeed a quark loop with
$k$ gluon vertices is ${\cal O}(N_f/N_c^{k/2})$.

This idea is supported by lattice simulations: apart from a rescaling due
to the difference in the beta function quenched simulations (no quarks
loops) differ from full QCD simulations tipically by $10\%$.

The $N_c\to \infty$ idea has lead to the solution of the $U(1)$
problem\cite{r9,r9a}.
Although it is difficult to perform calculations in the limt, a number of
very important qualitative results can be obtained in this approach, which
go deeply into the structure of the theory.

A number of non perturbative models exist of QCD which  assume that
some amplitudes are dominant and provide a good approximation to the full
theory.

Among them we quote the instanton gas (or liquid) model\cite{r10}, and the
stochastic vacuum model\cite{r11}.
\item Instanton gas (liquid) model. Perturbation theory (sect. 1.3) can
be viewed as a saddle point approximation to the Feynman integral,
approximating it with a gaussian expansion around the trivial saddle
point with zero fields. For memory the saddle point approximation to an
integral is
\begin{equation}
\int dx g(x) e^{i f(x)} =
\int d\delta g(\bar x+\delta) e^{i[f(\bar x) + \frac{f''}{2}\delta^2]}
\left[\sum_n\frac{\displaystyle (i \Delta f)^n}{n!}\right]
\label{eq1.28}
\end{equation}
with
\[\Delta f = f(\bar x +\delta)  - f(\bar x) -
\frac{f''(\bar x)}{2}\delta^2\]
and $\bar x$ a saddle point $f'(\bar x) = 0$. It reduces  to a gaussian
integral if $g$ and $\Delta f$ are expanded in powers of $\delta$.

In principle all saddle points should be included, and their contribution
added to get a better approximation.

For the Feynman integral eq.(\ref{eq1.28}) reads
\begin{equation}
Z = \int[dq]e^{-S(q)}\simeq
\int [d\delta] {\rm exp}\left[-\bar S - \frac{1}{2}\frac{\displaystyle
\delta^2 \bar S}{\displaystyle \delta q_i\delta q_j}\delta_i\delta_j
\right]\sum\frac{(-1)^n}{n!}(\Delta S)^n
\label{eq1.29}
\end{equation}
where
\[ q = \bar q + \delta\qquad
\left.\frac{\delta S}{\delta q}\right|_{q=\bar q} \hskip-5pt = 0
\qquad \bar S = S(\bar q)\qquad
\Delta S = S[\bar q+ \delta] - \bar S -
\frac{1}{2}\frac{\displaystyle
\delta^2 \bar S}{\displaystyle \delta q_i\delta q_j}\delta_i\delta_j
\]
Extending the approximation to all saddle points means to go beyond
perturbation theory: all the solutions $\bar q$ of the classical
equations of motion $\left.\frac{\delta S}{\delta q}\right|_{q=\bar q}
\ = 0$ have to be included, which have finite action $\bar S$. If $\bar
S=\infty$ the factor $e^{-S}$ in (\ref{eq1.29}) will kill the
contribution.

Classical solutions with finite euclidean action are called instantons.
they have non trivial topology. Indeed finite action means finite $\int
F^2 d^4 x$. To have convergence
\[ F^2\mathop{\simeq}_{|x|\to \infty} {\cal O}\left(\frac{1}{r^4}
\right)\]
As a consequence $A_\mu$ is a pure gauge on the sphere $S_3$ at infinity,
resulting in a mapping of $S_3$ on the gauge group. The jacobian of this
mapping is given by
\[ \frac{d U}{d\sigma^\mu}  = k_\mu \]
where $d U$ is the volume element of the group,
normalized to $\int d U = 1$,
$d\sigma^\mu$ the element
of the surface. The chern number
\[\int_{S_3} k_\mu d\sigma^\mu = n\]
is an integer (topological charge) and counts how many times the group is
swept in the mapping. By Stokes theorem
\[ \int_{S_3} k_\mu d\sigma_\mu = \int (\partial^\mu k_\mu) dV = n\]
$Q(x) = \partial^\mu k_\mu$ is the density of topological charge. In QCD
\begin{equation}
Q(x) = \frac{g^2}{32\pi^2} G^a_{\mu\nu}{\tilde G}^a_{\mu\nu}
\label{eq1.30}
\end{equation}
where
${\tilde G}^a_{\mu\nu} = \frac{1}{2}\varepsilon{\mu\nu\rho\sigma}
G^a_{\rho\sigma}$ is the dual of $G^a_{\mu\nu}$.

Topology ensures stability of the classical solutions.

Explicit solutions (B.P.S.T. instantons) were first found in
ref.\cite{r12}.

A line of research attempting to classify all possible instantons
developed immediately after the appearance of ref.\cite{r12}. A model was
strongly pushed assuming that the Feynman integral is dominated by quasi
solutions, with instantons at distances large enough that they can be
considered as independent (instanton gas)\cite{r13}.
The attempt was
frustrated by infrared problems, and by the failure to describe
confinement. The model was subsequently replaced by allowing overlapping
between instantons (instanton liquid), which is useful to describe chiral
properties of the theory\cite{r10}.
\item Stochastic vacuum. The idea is to compute physical quantities in
terms of gauge invariant field strength correlators. The simplest of them
is
\begin{equation}
{\cal D}_{\mu\nu,\rho\sigma}
=\langle G_{\mu\nu}(x) S^{adj}_C(x,0) G_{\rho\sigma}(0)\rangle
\label{eq1.31}
\end{equation}
whith $S^{adj}_C$ the parallel transport in the adjoint representation
from
$0$ to $x$ along the path $C$.
A cluster expansion is then performed, and higher irreducible correlators
are neglected.

Correlators extracted from lattice simulations\cite{r14} are used as an
input.

A good phenomenological model to describe high energy scattering between
hadrons and heavy quarks physics\cite{r15}.
\end{enumerate}
\section{Non perturbative effects in QCD. Vacuum condensates.}
\subsection{The SVZ sum rules.}
The vacuum condensates were first introduced in ref.\cite{r7} to
parametrize non perturbative effects.

Consider the Wilson operator product expansion (OPE) of the product of
two e.m. currents
\begin{equation}
T\left( j^\mu(x) j^\nu(0)\right)
\mathop{\sim}_{|x|\to 0}
C^{\mu\nu}_I(x) \cdot I +
C^{\mu\nu}_G(x) \frac{\beta(g)}{g} G^a_{\rho\sigma}G^a_{\rho\sigma}
+ C^{\mu\nu}_{\bar\psi\psi}(x) m\bar\psi\psi
+\ldots
\label{eq2.1}
\end{equation}
The product is written as a sum of local operators, ordered by increasing
dimension in mass, multiplied by coefficient functions.

Wilson OPE is a theorem in perturbation theory, and is presumably valid
in QCD at short distances.

In Fourier transform, taking the vacuum expectation value ({\em vev}) and
renormalizing one gets
\begin{equation}
\Pi_{\mu\nu}(q) \equiv
\int d^4x e^{iqx}\langle0|T\left( j_\mu(x) j_\nu(0)\right)|0\rangle =
(q_\mu q_\nu - q^2 g_{\mu\nu}) \tilde \Pi(q^2)\label{eq2.2}
\end{equation}
Here $\tilde \Pi(q^2) = \Pi(q^2) - \Pi(0)$ and, from Eq.(\ref{eq2.1})
\begin{equation}
\tilde \Pi(q^2) = C_1 \langle 0| I|0\rangle +
C_G\frac{G_2}{q^4} + C_{\bar\psi\psi}\frac{G_\psi}{q^4}
\label{eq2.3}
\end{equation}
$C_1, C_G, C_{\bar\psi\psi}$ are dimensionless coefficients, which are to
be computed in perturbation theory and $ \langle 0| I|0\rangle = 1$,
\begin{equation}
G_2 \equiv \langle 0| \frac{\beta(g)}{g} : G^a_{\mu\nu}
G^a_{\mu\nu}|0\rangle\label{eq2.4}
\end{equation}
has dimension 4, is the {\em vev} of the dilatation anomaly and is
known as gluon condensate.
\begin{equation}
G_{\psi} = \langle 0| m :\bar\psi \psi :|0\rangle
\label{eq2.5}
\end{equation}
has also dimension 4, is related to the spontaneous breaking of chiral
symmetry, and is known as quark condensate.

$G_2$ and $G_\psi$ are not defined in perturbation theory, where are
usually put equal to zero, so that only the first term survives in
eq.(\ref{eq2.3}). They parametrize non perturbative effects.

Eq.(\ref{eq2.3}) is the basis of SVZ sum rules.

The l.h.s. is parametrized by a dispersion representation
\begin{equation}
\tilde\Pi(q^2) = - q^2\int_{4 m_\pi^2}^\infty \frac{ds}{s}
\frac{\displaystyle R(s)}{\displaystyle s - q^2 + i\varepsilon}
\label{eq2.6}\end{equation}
where
\begin{equation}
R(s) =
\frac{\displaystyle \sigma_{e^+e^-\to h}(s)}{\displaystyle
\sigma_{e^+e^-\to \mu^+\mu^-}(s)}
\label{eq2.7}
\end{equation}
In perturbation theory $R(s)$ is a constant, modulo log's, $\sigma
\propto 1/s$, there is no scale, and in the r.h.s. of eq.(\ref{eq2.3})
only $C_1$ is present. In nature scale invariance is broken by resonances:
in the r.h.s. this reflects in the presence of condensates.

An appropriate weighting of both sides of the equation (34)
\begin{equation}
\int_{4 m_\pi^2}^\infty \frac{ds}{s}
\frac{\displaystyle R(s)}{\displaystyle s - q^2 + i\varepsilon}
\simeq
C_1  +
C_G\frac{G_2}{q^4} + C_{\bar\psi\psi}\frac{G_\psi}{q^4} \label{eq2.8}
\end{equation}
which emphasizes the ranges of $q^2$ in which both descriptions are
valid, relates the parameters of the resonances to the condensates. A
successfull phenomenology emerges, and a determination of
$G_2, G_\psi$\cite{r17}
\begin{equation}
G_2 = (.024\pm.011)\,{\rm GeV}^4\qquad \langle\bar\psi\psi\rangle = - .13
{\rm GeV}^3 \label{eq2.9}
\end{equation}
In fact the calculation of $C_1$ is affected by the presence of
renormalons, an arbitrariness which mimics the condensates, which, as a
consequence, are not well defined.

In spite of that sum rules work and give a consistent determination of
the condensates, independent of the choice of the currents considered.
\subsection{Determination of $G_2$ on the lattice.}
Consider the gauge invariant field strength correlators
\begin{equation}
{\cal D}_{\mu\nu,\rho\sigma}(x) =
\langle0| {\rm Tr}\left[G_{\mu\nu}(x) S_C(x,0) G_{\rho\sigma}(0)
S^\dagger_C(x,0)\right] |0\rangle \label{eq2.10}
\end{equation}
where
\[ G_{\mu\nu} = \sum^a t^a G^a_{\mu\nu}\qquad
A_\mu = \sum^a t^a A^a_\mu\]
with $T^a$ the generators in the fundamental representation and $S_C$ a
Schwinger parallel transport from $0$ to $x$ along the path $C$
\begin{equation}
S_C(x,0) = P\,{\rm exp}\left(i \int_{0,C}^x A_\mu(x) dx^\mu\right)
\label{eq2.11}
\end{equation}
In general the definition depends on the choice on the path $C$. This
will not be relevant for the following. For $C$ we shall assume a straight
line.
${\cal D}_{\mu\nu,\rho\sigma}(x) $ can be viewed as a split point
regulator \`a la Schwinger of $G_2$. The OPE at small $x$ gives
\begin{equation}
{\cal D}_{\mu\nu,\mu\nu}(x)\simeq \frac{c}{x^4} \langle I\rangle
+ c'\langle G_2\rangle +\ldots \label{eq2.12}
\end{equation}
with $c,c'$ computable coefficients in perturbation theory.
$\langle G_2\rangle$ is defined if $c,c'$ can be unambigously computed.
We will see that this is not the case. A generic parametrization dictated
by symmetry is\cite{r11}
\begin{eqnarray}
{\cal D}_{\mu\rho,\nu\sigma}(x) &=&
(g_{\mu\nu} g_{\rho\sigma} - g_{\mu\sigma} g_{\nu\rho})[ D(x^2) +
D_1(x^2)]\label{eq2.13}\\
&&+ \left(x_\mu x_\nu g_{\rho\sigma} + x_\rho x_\sigma g_{\mu\nu}
- x_\mu x_\sigma g_{\nu\rho} -x_\nu x_\rho g_{\mu\sigma}\right)
\frac{\partial D_1}{\partial x^2}\nonumber
\end{eqnarray}
Choose $x^\mu$ along 0 axis. Then\cite{r14}
\begin{eqnarray}
{\cal D}_{||}(x^2) &\equiv& \frac{1}{3} {\cal D}_{0i} {\cal D}_{0i}
= {\cal D} + {\cal D}_1 + x^2 \frac{\partial D_1}{\partial x^2}
\label{eq2.14}\\
{\cal D}_{\perp}(x^2) &\equiv& \frac{1}{3}\sum_{i<j}^3
{\cal D}_{ij}{\cal D}_{ij} = {\cal D} + {\cal D}_1\nonumber
\end{eqnarray}
On the lattice $G_{\mu\nu}$ is a plaquette, the parallel transport is
provided by the links and the lattice version of ${\cal
D}^L_{\mu\nu,\rho\sigma}$ of the correlator is given in pictures by

\begin{minipage}{0.6\linewidth}
${\cal D}^L_{\mu\nu\rho\sigma}=\left\langle
\hbox{
\epsfxsize0.4\linewidth
\epsfbox{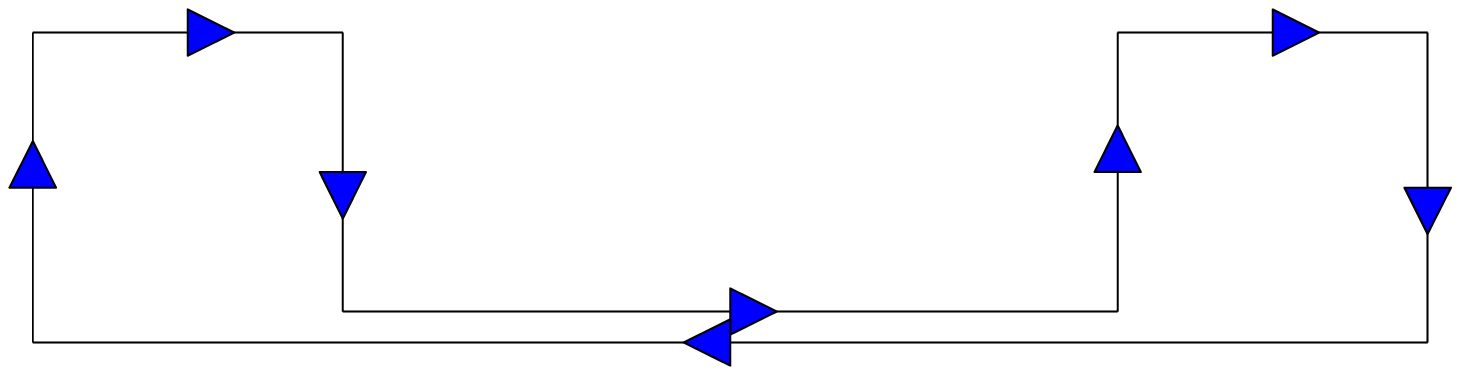}
}
\right.
$
\end{minipage}
\begin{minipage}{0.39\linewidth}\hskip-30pt
$-\frac{1}{N_c}\left.
\hbox{
\epsfxsize0.5\linewidth
\epsfbox{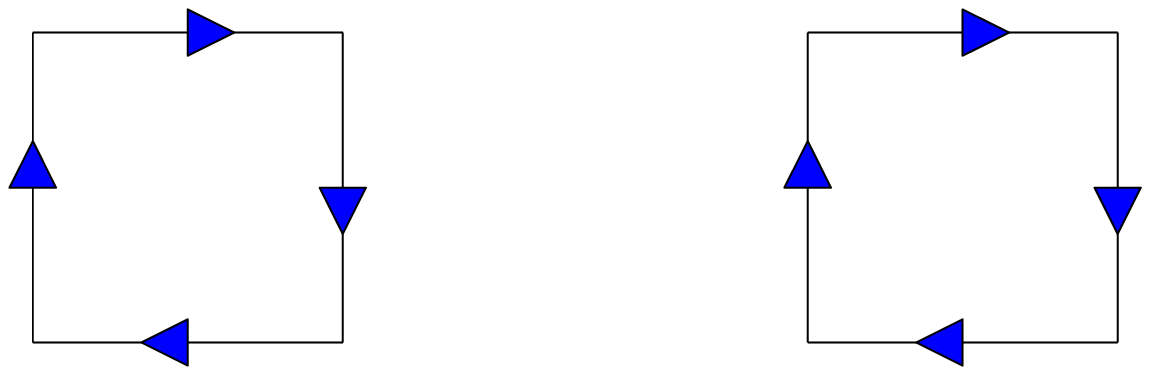}}\right\rangle$
\end{minipage}\par\noindent
\begin{minipage}{\linewidth}
\vskip-10pt\hskip70pt$\Pi_{\mu\nu}$
\hskip40pt$\Pi_{\rho\sigma}$
\hskip80pt$\Pi_{\mu\nu}$
\hskip15pt$\Pi_{\rho\sigma}$
\end{minipage}

The term subtracted insures that the quantum number exchanged is in the
adjoint representation.

At large $\beta = 2 N_c/g^2$ (see sect. 4)
\begin{equation}
{\cal D}^L_{\mu\nu,\rho\sigma} =
a^4 {\cal D}_{\mu\nu,\rho\sigma} + {\cal O}(a^6) \label{eq2.16}
\end{equation}
${\cal D}^L_{||}$ and ${\cal D}^L_{\perp}$ for quenched $SU(3)$ are shown
in fig.1

They can be parametrized as
\[ {\cal D}_i^L= \frac{A_i}{x^4}e^{-x/\lambda'} + c G_2e^{-x/\lambda}
\]
Fig.2 shows the
behaviour in full QCD, 4 quark species.

The corresponding results for $G_2$ and $\lambda$ are
\begin{eqnarray}
SU(2)\;  {\hbox{quenched}}\cite{r19} & G_2 = (0.33\pm0.01) {\rm GeV}^4
\phantom{00} & \lambda = .16\pm.02\,{\rm fm}
\label{eq2.17}\\
SU(3) \;   {\hbox{quenched}}\cite{r14} & G_2 = (0.15\pm0.03) {\rm
GeV}^4\phantom{00} &
\lambda = .22\pm.02\,{\rm fm}
\label{eq2.18}\\
SU(3) \;   {\hbox{full QCD}}\cite{r20} & G_2 = (0.022\pm0.005) {\rm GeV}^4
&
\lambda = .34 {\rm fm}
\label{eq2.19}
\end{eqnarray}
The value obtained from SVZ sum rules is\cite{r17}
\[ G_2 = (0.024\pm0.011) {\rm GeV}^4\]
in excellent agreement with eq.(\ref{eq2.19}).

However, like in sum rules the determination of the coefficients $c$ in
eq.(\ref{eq2.12}) is again ambigous by terms which mimic the second term
in $\langle G_2\rangle$, due to renormalons. Nevertheless everything
works and compares satisfactorily with sum rules.

An alternative determination can be obtained by measuring the expectation
value of the plaquette (density of action). Indeed\cite{r21}
\[ 1 - \Pi_{\mu\nu}\propto g^2 G^a_{\mu\nu}G^a_{\mu\nu} a^4\]
so that
\[ \langle0|1 - \Pi_{\mu\nu}|0\rangle \propto G_2 a^4\]
Here again a term exists which scales like $a^4$ (eq.({\ref{eq1.26})),
but a perturbative contribution, corresponding in the continuum to a
quadratic divergence is present, which must be subtracted. This term is
again ambigous by contributions which mimic $G_2$.

Also here, however, the determination is in agreement with (\ref{eq2.17}),
(\ref{eq2.18}).

\subsection{Lattice determination of the chiral condensate
$\langle\bar\psi\psi\rangle$}
The gauge invariant quark correlator
\begin{equation}
S(x) = \langle \bar\psi(x) S(x,0) \psi(0)\rangle
\label{eq2.20}
\end{equation}
can also be determined on the lattice\cite{r22}.

In the continuum limit
\begin{equation}
S(x) \mathop\simeq_{|x|\to 0} \frac{c}{x^2} + c_1
\langle\bar\psi\psi\rangle\label{eq2.21}
\end{equation}
On the lattice
\begin{equation}
S^L(x) = \frac{B a^3}{x^2} + a^3 \langle\bar\psi\psi\rangle
e^{-x/\lambda_f} + {\cal O}(a^5)\label{eq2.22a}
\end{equation}
The result of the fit gives
\begin{eqnarray}
|\langle\bar\psi\psi|^{1/3} &=&\left\{
\begin{array}{lll}
(205\pm21){\rm MeV} & am = 0.02 & \frac{m_\pi}{m_\rho} = 0.65(3)\\
(160\pm12){\rm MeV} & am = 0.01 & \frac{m_\pi}{m_\rho} = 0.57(2)
\end{array}\right.\nonumber\\
\lambda &=& 1.4 m_\pi^{-1}
\label{eq2.22}
\end{eqnarray}
A more careful analysis of the chiral limit is needed.

In conclusion the condensates can be extracted from first principle on
the lattice, and agree with the phenomenological determinations, in spite
of the fact that $G_2$ is undefined by renormalon ambiguities both in the
lattice determination and in the sum rules.

The lattice correlators provide the gluonic correlation length. In
quenched QCD they are smaller by a factor 3 with respect to
full QCD.

The lattice correlators are used as an input in the stochastic vacuum
model.

\section{Confinement of colour.}
\subsection{Introduction.}
Quarks and gluons have never been observed: asymptotic states in QCD are
colour singlets. This phenomenon is known as confinement of colour.

In nature the ratio of the abundance of quarks $n_q$ compared to that of
nucleons $n_p$ has an upper limit\cite{r23}
\begin{equation}
\frac{n_q}{n_p} < 10^{-27}\label{eq4.1}
\end{equation}
corresponding to Millikan analysis of $\sim 1$~g of matter, in search of
fractional charges.

In the absence of confinement the Standard Cosmological Model predicts
$n_q/n_p\simeq 10^{-12}$. The argument is as follows\cite{r24}. Let
$\sigma_0 =
\lim_{v\to 0}v \sigma$, $\sigma$ being the cross section for the process
of hadronization in quark gluon plasma $q\,\bar q \to {\rm hadrons}$,
$q\,q\to \bar q+ {\rm hadrons}$. Quarks will decouple in the process of
cooling of the universe when the rate of burning is of the order of the
rate of expansion
\[ \sigma_0 n_q = G_N^{1/2} T^2\]
since $n_\gamma = T^3$, this gives
\[\frac{n_q}{n_\gamma} = \frac{G_N^{1/2}}{\sigma_0 T^2} =
\frac{10^{-19}}{\sigma_0 m_p T}\]
An estimate for $\sigma_0$ is $\sigma_0\leq m_\pi^{-2}$. A conservative
value for $T$ is $T\sim 10$~GeV and this give
\[ \frac{n_q}{n_\gamma}\geq 10^{-21}\]
or, since $n_p/n_\gamma \simeq 10^{-9}$, $n_q/n_p\geq 10^{-12}$.

A factor $10^{-15}$ between expectation and the empirical upper limit is
too large to be explained by tuning small parameters. It can only be
explained in terms of symmetry, like e.g. the smallness of resistivity in
a superconductor\cite{r24a}.

The deconfined phase is observed on the lattice, by simulating QCD at
high temperature.

The partition function is obtained by computing the Euclidean Feynman
integal at infinit spatial extension, on a time interval $1/T$, $T$ being
the temperature.

This is done on a lattice $N_T\times N_S^3$ with $N_S\gg N_T$ and,in
terms of the lattice spacing $a$
\[ \frac{1}{T} = N_T a\]
In physical units (eq.({\ref{eq1.26}))
\begin{equation}
\frac{T}{\Lambda} = \frac{1}{N_T}{\rm exp}(b_0\beta)\label{eq4.3}
\end{equation}
Due to asymptotic freedom high temperature corresponds to high $\beta$,
i.e. to weak coupling ($\beta = 2N_c/g^2$), low temperature to strong
coupling.

The deconfining transition is detected by measuring\cite{r24b} either the
string tension $\sigma$ or the expectation value of the Polyakov line
$\langle  L \rangle$, which is the trace of the parallel transport from
$-\infty$ to
$+\infty$ along time axis, closing trough periodic boundary conditions.

The string tension is the constant appearing in the linear static
potential responsible for confinement,
\begin{equation}
V(r) = \sigma r\label{eq4.4}
\end{equation}
has dimension $m^2$, is related to the slope of the Regge trajectories
$S$ ($\sim 1$~GeV$^2$) by the relation
\begin{equation}
\sigma = \frac{1}{2\pi} S\label{eq4.5}
\end{equation}
$\sigma\neq 0$ means confinement.

The Polyakov line
\begin{equation}
\langle L\rangle = \langle {\rm Tr}\left\{ P{\rm exp}\left[
\int_{-{N_T a}}^{N_Ta} A_0(\vec x,t) dt\right]\right\}
\label{eq4.6}
\end{equation}
vanishes if $Z_N$ is a symmetry. This happens in the confined phase.

In the confined phase $\langle L\rangle $, which can be interpreted as
${\rm exp}(-\mu_q)$, $\mu_q$ being the chemical potential of a quark, has
zero expectation value since an infinite energy is required to add a free
quark to the system (confinement).

In the language of statistical mechanics a parameter which is different
from zero in the weak coupling regime, like $\langle  L \rangle$, is
called an order parameter, while a parameter like $\sigma$ which is non
zero in the strong coupling regime is called a disorder parameter.

The meaning of these denominations will be clear below, when the concept
of duality will be introduced.

A drawback of both $\sigma$ and $\langle  L \rangle$ is that they only
work for pure gauge theories in the absence of quarks. Indeed if dynamical
quarks are present the string tension is not defined, since taking $q\bar
q$ apart does not produce a potential energy but $q\bar q$ pairs. With
dynamical quarks $Z_N$ is not a simmetry any more, since quarks belong to
the fundamental representation, which has non zero triality, and cannot
be an order parameter. A true order disorder transition is related to
chiral symmetry, the corresponding disorder parameter is
$\langle\bar\psi\psi\rangle$, but it is not precisely understood what
chirality has to do with confinement.

A genuine order (or disorder) parameter for confinement is needed, which
can work both with and without quarks.

Indeed, in the spirit of $N_c\to\infty$, as described in sect.1.5,
the underlying dynamics must be the same, independent of $N_c$
and $N_f$, for sufficiently small $N_f$, apart from small corrections
${\cal O}(1/N_c)$.

This will be one of our main prejudices in the search of a mechanism for
confinement. The other prejudice will be that a suppression by a factor
$10^{-15}$ of $n_q/n_p$ can only be explained in terms of symmetry.

The symmetry of the confined (disorderd) phase can be understood in terms
of duality\cite{r25,r26}.

Duality is a deep concept in statistical mechanics and in field theory. It
applies to systems which admit non local excitations with non
trivial topology. The idea is that, besides the usual description in terms
of local fields appropriate to the weak coupling regime, a dual
description is possible, in which topological excitations become local,
and local fields become topological excitations.
In the dual description the original strong coupling regime becomes weak
coupling.

We shall illustrate the concept of duality on a simple system, the 2d
Ising model, in the next section.

As for QCD if the dual were known, i.e. if the non trivial topological
excitations were identified, which interact weakly in the confined phase,
an effective Lagrangian could be written, and the theory would be
essentially solved.

This already happens in $N=2$ supersymmetric version of the
theory\cite{r27}.

A big step forward has been done in that direction in ordinary QCD,
on the line of understanding the symmetry of the dual
excitations\cite{r28}. The results will be presented in sect.3.3
\subsection{Duality: the 2d Ising model.}
To illustrate the strategy used in QCD and the meaning of duality, we
shall describe in some detail the 2d Ising model, which is the prototype
system showing dual behaviour\cite{r26}, We will do that by a tecnique
allowing to explicitely compute the disorder parameter interms of spin
variables\cite{r29}.

The two dimensional Ising model is defined on a square lattice, by a
field variable
\begin{equation}
\sigma(\vec n) = \pm 1\label{eqs0}
\end{equation}
The partition function is
\begin{equation}
Z = \exp[-S[\sigma]/T]\label{eqs4}
\end{equation}
with
\begin{equation}
S[\sigma] = J\sum_{\vec n,\hat\mu}\sigma(\vec n)\sigma(\vec n+\hat\mu)
\label{eqs5}
\end{equation}
$J>0$, $\vec n$ is the site and $\hat\mu$ are the two independent vectors
joining a site with the neighbouring sites.

The model is exactly solvable, and presents two phases separated by a
critical temperature $T_c$. Defining the magnetization as
\[\langle\sigma\rangle = \lim_{V\to\infty}\frac{1}{V}\sum_{\vec n}
\langle\sigma(\vec n)\rangle\]
\[
\begin{array}{lcl}
\hbox{for } T < T_c & \langle\sigma\rangle\neq 0 & \hbox{(ordered phase)}
\\
\hbox{for } T > T_c & \langle\sigma\rangle = 0 & \hbox{(disordered phase)}
\end{array}\]
The critical temperature is related to the scale $J$ of the model by the
equation
\begin{equation}
T_c = \frac{2 J}{\ln(\sqrt{2}+1)}\label{eqs6}
\end{equation}
As $T$ approaches the critical temperature $T_c$ $\langle\sigma\rangle$
tends to zero as
\begin{equation}
\langle\sigma\rangle \mathop\simeq_{T\to T_c} (T_c- T)^\beta\qquad
\beta = \frac{1}{8}\label{eqs7}
\end{equation}
The correlation length $\xi(T)$ defined by the relation
\begin{equation}
\langle\sigma(\vec n)\sigma(\vec m)\rangle \mathop\simeq_{
|\vec n-\vec m|\to\infty} \;\exp\left[-\frac{|\vec n-\vec
m|}{\xi(T)}\right] + \langle\sigma\rangle^2 \label{eqs8}
\end{equation}
behaves as
\begin{equation}
\xi(T) \mathop\simeq_{T\to T_c} (T-T_c)^{-\nu}\qquad \nu = 1
\label{eqs9}
\end{equation}
A field theory in $1+1$ dimensions is defined at the critical point,
where the correlation length diverges (eq.(\ref{eqs9})), and the coarse
structure of the lattice becomes irrelevant.

The action can be written, up to an irrelevant constant as
\begin{equation}
S = \frac{J}{2} \sum_{\vec n,\hat\mu}\left[\Delta_\mu\sigma(\vec n)
\right]^2\label{eqs10}
\end{equation}
with
\begin{equation}
\Delta_\mu\sigma(\vec n) = \sigma(\vec n+\hat \mu) - \sigma(\vec n)
\label{eqs11}
\end{equation}
The equation of motion reads
\begin{equation}
\Delta_\mu\Delta_\mu\sigma(\vec n) = 0
\label{eqs12}
\end{equation}
and the current
\begin{equation}
j_\mu = \frac{1}{2}\varepsilon_{\mu\nu}\Delta_\mu\Delta_\nu\sigma
\label{eqs13}
\end{equation}
is identically conserved
\begin{equation}
\Delta_\mu j_\mu = 0 \label{eqs14}
\end{equation}
The correspnding constant of motion is
\begin{equation}
Q = \int dx\,j_0(t,x) = \sigma(x=+\infty) - \sigma(x=-\infty)
\label{eqs15a}
\end{equation}
$Q$ is the number of kinks minus the number of antikinks, and has a
topological meaning.

The dual description\cite{r26} of the system is constructed as follows. A
dual lattice is defined by associating a site of it to each link of the
original lattice and a  dual variable $\sigma^*(j)$ on the dual lattice
via its correlation functions
\begin{equation}
\langle\sigma^*(\vec i)\sigma^*(\vec j)\rangle\equiv
\frac{\tilde Z}{Z}\label{eqs15}
\end{equation}
$\tilde Z$ is obtained from $Z$, eq.(\ref{eqs4}) by changing the sign of
the links $[J\sigma(\vec n)\sigma(\vec n+\hat\mu)]$ along an arbitrary
path in the dual lattice joing $\vec i$ to $\vec j$. Then\cite{r26}:
\begin{itemize}
\item[(i)] $\sigma^*$ is a dicotomic variable like $\sigma$,
$\sigma^*(\vec i) =\pm 1$.
\item[(ii)] $\tilde Z$ is independent of the choice of the path from $\vec
i$ to
$\vec j$ on the dual lattice.
\item[(iii)] In the thermodynamic limit $(V\to\infty)$
\begin{equation}
Z[\sigma,T] \equiv Z[\sigma^*,T^*]\label{eqs16}
\end{equation}
with
\begin{equation}
{\rm sinh}\frac{2}{T} = \frac{1}{{\rm sinh}\frac{2}{T^*} }
\label{eqs17}
\end{equation}
or
\begin{equation}
T \sim \frac{1}{T^*}
\label{eqs18}
\end{equation}
\end{itemize}
The partition function of the new variables has the same dependence on
$\sigma^*$ as the original partition function had on $\sigma$, except that
high temperature (strong coupling) is mapped into low temperature and
viceversa.
\[ \langle\sigma^*\rangle = 0\quad T< T_c\qquad\quad
\langle\sigma^*\rangle \neq 0\quad T > T_c \]
or
\begin{equation}
\langle\sigma\rangle\langle\sigma^*\rangle = 0
\label{eqs19}
\end{equation}
$\langle\sigma^*\rangle$ is the order parameter in
the dual description and is called a disorder parameter. Strong coupling
in terms of $\sigma$ becomes weak coupling for $\sigma^*$ and viceversa.

An explicit construction of $\langle\sigma^*\rangle$ can be done which
evidences its meaning of creation operator of a kink\cite{r29}. As a
consequence $\langle\sigma^*\rangle\neq 0$ is nothing but condensation of
kinks, or spontaneous breaking of the topological symmetry
(\ref{eqs14}). In the hot phase the vacuum is a superposition of states
with different values of $Q$, the number of kinks.

The disordered phase looks ordered in the dual description.

The operator which creates a kink at $x=x_0$ and time $t_0$ can be
written\cite{r29}
\begin{equation}
\mu^{(t)} =\exp\left[\frac{2 J}{T} \sum_{x\geq x_0}\Delta_0(x,t_0)
\right]\label{eqs20}
\end{equation}
$\Delta_0(x,t_0)$ is nothing but the conjugate momentum to $\sigma$ with
the Hamiltonian  eq.(\ref{eqs10}) and (\ref{eqs20}) is nothing but the
version for a discrete euclidean field of the well known relation
\begin{equation}
e^{ipa}| x\rangle = | x + a\rangle\label{eqs21}
\end{equation}
by which the position is shifted.

In field theory the analog of $x$ is the field $\Phi$, the analog of $p$
its conjugate momentum, and eq.(\ref{eqs21}) is the tool to add a
classical topological excitation to the field configuration.
Some care is necessary when the field is compact\cite{r28} or even
discrete.

The expectation value of $\mu$ can be computed numerically: a more
convenient technique is to
compute the quantity
\begin{equation}
\rho = \frac{d}{d\beta}\ln\langle\mu\rangle\label{eqs22}
\end{equation}
\begin{equation}
\langle\mu\rangle = \exp\left[\int_0^\beta \rho(\beta')d
\beta'\right]\label{eqs23}
\end{equation}
The phase transition to the ordered phase produces a sharp drop in
$\langle\mu\rangle$, or a negative peak on $\rho$.

$\rho$ vs $T$ is shown in fig.3 for different sizes of the lattice. At
large $\beta$ $\rho$ can be computed in perturbation theory, and the
comparison with data is shown in the figure. As $L\to \infty$, $\rho\to
-\infty$, or $\langle\mu\rangle\to 0$, as expected. At small $\beta$
$\rho$ is a constant compatible with zero, and lattice size independent,
corresponding to $\langle\mu\rangle = 1$. In the transition region where
$\langle\mu\rangle \sim (T_c-T)^\delta$ a finite size scaling analysis can
be performed to extrapolate to the thermodynamic limit
\begin{equation}
\langle \mu\rangle = (\beta_c -\beta)^\delta
f\left(\frac{L}{\xi},\frac{a}{\xi}\right)\label{eqs24}
\end{equation}
If $\frac{a}{\xi}\ll 1$, $f$ only depends on $L/\xi = L (1-T/T_c)^\nu$,
and
\begin{equation}
\langle\mu\rangle \sim (\beta-\beta_c)^\delta\,f\left[L^{1/\nu}(
\beta_c - \beta)\right]\label{eqs25}
\end{equation}
or
\begin{equation}
\frac{\rho}{L^{1/\nu}} = \Phi(L^{1/\nu}(\beta_c - \beta))
\label{eqs26}
\end{equation}
This relation allows to determine both $\nu$ and $\beta_c-\beta$ from
data coming from different lattice sizes.

The result for scaling is shown in fig.4. The result is compatible with
the known values
\[ \nu = 1\qquad \beta_c = 0.44068\]
and gives $\delta = 0.120(5)$ to be compared with the exact value
$\delta_{ex} = 0.125$.

The construction of the disorder parameter by the rule (\ref{eqs21}) has
also been successfull for other systems which are expected to have a dual
behaviour: the $xy$ model in 3d (liquid $He_4$\cite{r30}), the compact
$U(1)$ gauge theory in 4d\cite{r31}, the Heisenberg model\cite{r32}.

In each case the proper topological excitations condense in the
disordered phase.

We have then used the same method for QCD.

\section{Duality and confinement in QCD}
As explained in sect.3.1, we expect confinement to be produced by a
symmetry property of the vacuum, related to the condensation of
topological excitations. Natural topology of three dimensional
configurations comes from the mapping of the sphere at infinity, $S_2$, on
a group. The corresponding topological quantum number is magnetic charge.

Condensation of magnetic charges in the vacuum is the magnetic analog of
condensation of electric charges, which is known as superconductivity,
and is named ``dual superconductivity''\cite{r33}.

The picture fits with the original proposal of ref's.\cite{r34,r35}, which
ascribed confinement to the formation of dual Abrikosov lines, by the
chromoelectric field of a $q\bar q$ pair, giving an energy proportional
to the distance, or a string tension (eq.(\ref{eq4.4})).

Monopoles in a non abelian gauge theory exist as stable particles in the
Higgs phase of a theory coupled to a scalar field in the adjoint
representation\cite{r36,r37}. In the absence of Higgs field a conserved
magnetic charge can be defined anyhow by the procedure known as abelian
projection. We shall briefly summarize it for the simple case of $SU(2)$;
the extension to higher groups is routine\cite{r28}.

A gauge invariant field strength $F_{\mu\nu}$ can be defined\cite{r36} as
follows
\begin{equation}
F_{\mu\nu} = \hat\Phi\vec G_{\mu\nu} -
\frac{1}{g}\hat\Phi(D_\mu\hat\Phi\wedge D_\nu\hat\Phi)
\label{eqs27}
\end{equation}
here $\vec G_{\mu\nu}$ is the field strength tensor
\begin{equation}
\vec G_{\mu\nu} = \partial_\mu \vec A_\nu - \partial_\nu\vec A_\mu
+ g \vec A_\mu\wedge \vec A_\nu \label{eqs28}
\end{equation}
$\vec\Phi$ is any field belonging to the adjoint representation and
\begin{equation}
\vec \Phi = \frac{\vec\Phi}{|\vec\Phi|}\label{eqs29}
\end{equation}
is the unit vector of its direction in colour space. $\hat\Phi$ is
defined everywhere except where $\vec\Phi=0$.
\begin{equation}
D_\mu = \partial_\mu + g \vec A_\mu\wedge \label{eqs30}
\end{equation}
is the covariant derivative.

Both terms in (\ref{eqs27}) are separately gauge invariant.

The combination is chosen in such a way that terms in $A_\mu A_\nu$
cancel. It is easy to check that
\begin{equation}
F_{\mu\nu} = \hat\Phi(\partial_\mu\vec A_\nu - \partial_\nu\vec A_\mu)
-\frac{1}{g}\hat\Phi(\partial_\mu\hat\Phi\wedge \partial_\nu\hat\Phi)
\label{eqs30a}
\end{equation}
The dual of $F_{\mu\nu}$
\begin{equation}
F^*_{\mu\nu} = \frac{1}{2}\varepsilon_{\mu\nu\rho\sigma} F_{\rho\sigma}
\label{eqs31}
\end{equation}
defines a magnetic current
\begin{equation}
j_\nu = \partial_\mu F^*_{\mu\nu}\label{eqs32}
\end{equation}
which is identically conserved.

The corresponding magnetic charge is a constant in time. This is true for
any choice of the local operator $\hat\Phi$. There is an infinite number
of topological constants of the motion in QCD.
In the gauge in which $\hat\Phi$ is constant, e.g. oriented along the 3
axis, the second term in eq.(\ref{eqs30a}) drops and $F_{\mu\nu}$ becomes
an abelian field
\begin{equation}
F_{\mu\nu} = \partial_\mu A_\nu^3 - \partial_\nu A_\mu^3
\label{eqs33}
\end{equation}
The corresponding gauge transformation is called abelian projection. If
$\vec\Phi$ were a Higgs field, the gauge would be the unitary gauge.

It can be shown that\cite{r33} monopoles can exist at the sites where
$\vec\Phi=0$. $\vec\Phi=0$ describes world lines of monopoles.

Of course people like to think of monopoles as stable, observable
particles. This is not the case in general for the monopoles defined by
the abelian projection. However the corresponding magnetic symmetry
exists and is well defined. One can then investigate if the vacuum
is invariant under it, or it is a superposition of states with different
magnetic charges, like the Bogoliubov vacuum of superconductivity.

This will in any case give information on the symmetry pattern of the
confining mechanism.

As already noticed, there are infinitely many magnetic symmetries, one
for each abelian projection. There has been a tendency, during the years
to think that some abelian projection is better than others, and really
identifies not only a symmetry, but also the real excitations which would
be weakly coupled in the dual description of QCD\cite{r38}. The idea was
supported by the so called abelian dominance and monopole dominance. The
$U(1)$ degrees of freedom emerging
from this projection
and in special those related to magnetic charges
seem to account of a large fraction of observed
quantities (string tension, condensates\ldots).

An alternative idea\cite{r33} is that all abelian projections define
magnetic charges which are physically equivalent. This would mean that
the yet unknown excitations which become the fundamental fields in the
dual description of QCD, have non zero magnetic charge in all abelian
projections. If true this would be a very important step in understanding
the dual description of QCD. We have constructed\cite{r28} a disorder
parameter for the generic magnetic symmetry, defined by abelian
projection, by using the techniques illustrated in sect.3.2. We have then
checked its behaviour as a function of the temperature by lattice
simulations. We have done that for a number of different independent
choices of the field $\vec\Phi$ which defines the abelian projection,
both in $SU(2)$ and in $SU(3)$ pure gauge theory\cite{r28} and in full
QCD\cite{r39}.

The behaviour of $\rho$ across a deconfining transition is shown in
fig.5.

The finite size analysis (fig.6) allows to determine the critical index
$\delta$ of the disorder parameter, the critical index $\nu$ of the
correlation length and $\beta_c$. The latter two quantites are known by
independent method\cite{r40}. The results can be summarized as follows
\begin{itemize}
\item[(i)] The determinations of $\beta_c$ and $\nu$ are consistent with
other methods, showing that we are really observing the deconfining
transition.
\item[(ii)] Condensation exists for all the abelian projections.
\item[(iii)] The critical index $\delta$ is independent of the abelian
projection.
\end{itemize}
The definition of the disorder parameter is legitimate also in the
presence of quarks. There is a peak in $\rho$, analogous to that of
fig.5 at the chiral phase transition. Analysis is in progress to
determine the critical index $\nu$ and to check that $\beta_c$ coincides
with that of the chiral transition. Our disorder parameter describes
confinement also in the presence of quarks, and this is consistent
with the idea of
$N_c\to\infty$.
\section{Conclusions.}
Confinement is one of the most fascinating problems in theoretical
physics.

Lattice simulations have provided important hints to understand its
mechanism.

The transition confinement-deconfinement is for sure an order disorder
transition.

Duality seems at work, the confined phase breaks the magnetic symmetry
defined by all abelian projections. The dual excitations have magnetic
charge in all of them.

The dream is to identify the dual excitations, as was done in $N=4$ SUSY
QCD in ref.\cite{r27}. This would give the solution of QCD in terms of a
weakly interacting effective lagrangian of dual fields.

\section*{Acknowledgements}
Partially supported by EC TMR Program
ERBFMRX-CT97-0122, and by MURST, project: ``Fisica Teorica delle
Interazioni Fondamentali''.

\newpage
{\bf Figure Captions}
\vskip 0.2in
\par\noindent Fig.1 Field strength correlators in pure gauge $SU(3)$, 
ref.\cite{r14}.
\par\noindent Fig.2 Field strength correlators in full QCD, ref.\cite{r20}.
\par\noindent Fig.3 2d Ising model. $\rho$ vs $T$ for different lattice sizes.
\par\noindent Fig.4 2d Ising model. Finite size scaling, eq.(\ref{eqs26}).
\par\noindent Fig.5 $SU(3)$ gauge theory. $\rho$ vs $\beta$ for 
different abelian
projections. The peak signals the confinement.
\par\noindent Fig.6 $SU(3)$ gauge theory. Finite size scaling of $\rho$.
\end{document}